\documentclass[aip, apl, reprint, amsmath,amssymb,superscriptaddress]{revtex4-1}
\usepackage{graphicx}
\usepackage[colorlinks=true,citecolor=blue]{hyperref}
\usepackage[normalem]{ulem}
\usepackage{xcolor}
\usepackage{comment}
\usepackage{cancel}
\usepackage{enumitem}

\newcommand*{\citen}[1]{%
  \begingroup
    \romannumeral-`\x 
    \setcitestyle{numbers}%
    \cite{#1}%
  \endgroup   
}

\makeatletter
\def\@email#1#2{%
 \endgroup
 \patchcmd{\titleblock@produce}
  {\frontmatter@RRAPformat}
  {\frontmatter@RRAPformat{\produce@RRAP{*#1\href{mailto:#2}{#2}}}\frontmatter@RRAPformat}
  {}{}
}%
\makeatother

\begin{document}

\preprint{AIP/123-QED}

\title{Enabling high giant magnetoresistance in simple spin valves with ultrathin seed and free layers}
\author{Sachli Abdizadeh}
 \affiliation{Department of Physics, Virginia Tech, Blacksburg, Virginia 24061, USA}
\author{Rachel E. Maizel}
 \affiliation{Department of Physics, Virginia Tech, Blacksburg, Virginia 24061, USA}

\author{Dylan L. Haymore}
\affiliation{%
Academy of Integrated Science, Virginia Tech, Blacksburg, Virginia 24061, USA
}%

\author{Jing Zhao}
\affiliation{%
Department of Geosciences, Virginia Tech, Blacksburg, Virginia 24061, USA
}%

\author{F. Marc Michel}
\affiliation{%
Department of Geosciences, Virginia Tech, Blacksburg, Virginia 24061, USA
}%
\affiliation{%
Academy of Integrated Science, Virginia Tech, Blacksburg, Virginia 24061, USA
}%

\author{Satoru Emori}
 \email{sachli@vt.edu; semori@vt.edu}
\affiliation{%
Department of Physics, Virginia Tech, Blacksburg, Virginia 24061, USA
}%
\affiliation{%
Academy of Integrated Science, Virginia Tech, Blacksburg, Virginia 24061, USA
}%

\date{\today}

\begin{abstract}
Emerging spin-orbit-torque devices based on spin valves require a thin magnetic free layer to maximize the torque per moment.  However, reducing the free-layer thickness to $\lesssim 2$ nm deteriorates the giant magnetoresistance (GMR) signal for electrical readout. Here, we demonstrate that the addition of a 1-nm Cu seed layer promotes sharp interfaces in simple polycrystalline Co-based spin valves, enabling high GMR ratios of 5–7\% at sub-2-nm free-layer thicknesses. Our work offers a pathway for engineering high-signal GMR readout in spin-orbit-torque digital memories and neuromorphic computers.  
\end{abstract}

\maketitle

Spin valves, consisting of a ``free layer" with a switchable magnetization and a ``fixed layer" with a pinned magnetization, offer a promising platform for next-generation nanomagnetic devices driven by spin-orbit torques~\cite{RevModPhys.91.035004,9427163}. In this device architecture, an in-plane electric current through the fixed layer produces a spin current whose propagation and polarization directions are out-of-plane\cite{DAVIDSON2020126228,Kim2024}. This out-of-plane spin current is predicted by symmetry to exert an anti-damping torque on the free-layer magnetization, potentially enabling field-free perpendicular switching for digital memories\cite{Baek2018}. The torque from out-of-plane spin current could also drive large-angle precession\cite{PhysRevB.111.054425} in easy-to-fabricate spin valves without exquisite tuning of magnetic anisotropy\cite{Montoya2023,Divinskiy2019}, hence holding promise for oscillators in neuromorphic computing\cite{Grollier2020}. 
Furthermore, the giant magnetoresistance (GMR) of the spin valve allows for reading out the state of the spin-orbit torque device~\cite{Chen2020,Liu2020}.   

In a spin valve, the spin-orbit torque per unit moment scales inversely with the free-layer thickness~\cite{RevModPhys.91.035004,9427163}. Hence, a thinner free layer is necessary to enhance the driving efficiency of the device.  
At the same time, a magnetic layer thickness reduced to $\lesssim2$ nm ($\lesssim10$ atomic monolayers) typically leads to increased spin-flip scattering at interfaces, compounded by poor film quality, which decreases the current spin polarization\cite{Bass_2007,Haidar2013SpinPolThin,Erekhinsky2010Surface,Alcer2017Mesoscopic,Cormier2010DomainWalls} and hence the GMR response \cite{mi12091010,Elsafi_2025,parkin1990,PhysRevB.79.174421,parkin_oscillations_1990,Zsurzsa2022,Dieny1991}.
Studies over three decades have elevated GMR by growing magnetic multilayers on several-nm-thick seed layers of metal like Cu\cite{Parkin1991,Egelhoff1992CuBuffer,Kanai1993,489821,chihaya2004effect,bouziane2006buffer}, setting the desirable (111) texture of the subsequent face-centered cubic (FCC) magnetic layers. However, for emerging spin-orbit torque applications, it is essential to minimize the conductive Cu seed-layer thickness so that a larger fraction of the input charge current can flow in the spin valve's fixed layer.
It is not yet known whether a spin valve can maintain strong GMR with not only a thin free layer, but also a minimal Cu seed layer, e.g., down to $\approx$1-nm thickness that barely constitutes a continuous layer.


\begin{figure} 
    \centering
    \includegraphics[width=0.45\textwidth]{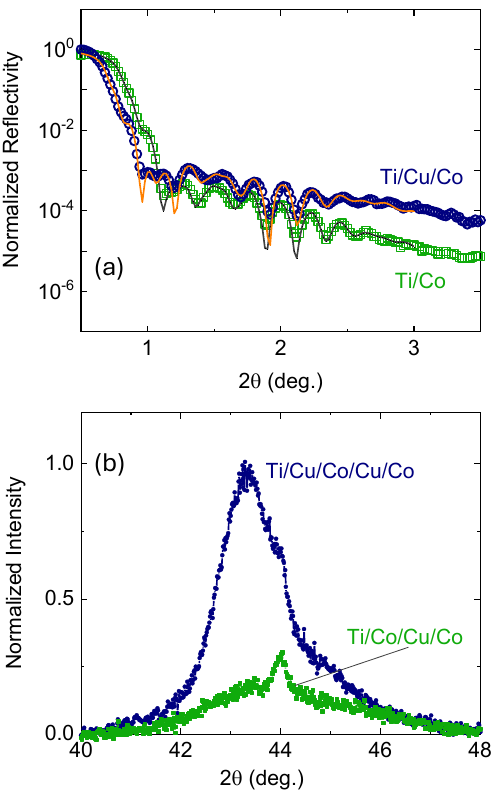} 
    \caption{\raggedright (a) XRR data (symbols) and fits (solid curves) for Ti- and Ti/Cu-seeded Co(30 nm) films. The fits were used to extract the interfacial width $\sigma$ (diffuseness or roughness). (b) XRD of Ti- and Ti/Cu-seeded Co(3.0 nm)/Cu(3.5 nm)/Co(3.0 nm) stacks. }
    \label{fig:fig1}
\end{figure}

Here, we demonstrate that a 1-nm Cu seed layer promotes sharp interfaces and pronounced texturing in the free layer, yielding GMR ratios of $\approx 5-7$\% with $\lesssim 2$-nm free layers. These values far exceed the ratios of $\approx 1-2$\% previously reported with $\lesssim2$-nm free layers\cite{Chan2016UltrathinSV,Romera2009UltrathinGd,Dieny1991} and are comparable to conventional spin valves with thicker free layers \cite{Ustinov2012HysteresisFree,Dieny1991,PhysRevLett.71.1641,Kim2019TailoringGMR}. 
Our findings present an effective approach toward robust GMR reading for next-generation spin-orbit-torque memories and neuromorphic computers.  

We focus on film stacks that can be readily sputter-grown without requiring post-annealing \cite{Hong2003} or stringent conditioning of the trace background gases \cite{Egelhoff1996}. The film stacks were grown on $\approx$ 5 mm $\times$ 5 mm pieces of thermally oxidized Si wafer (with 50-nm-thick SiO$_2$) by DC magnetron sputtering at an Ar sputtering gas pressure of 3 mTorr and a background pressure of $< 8 \times 10^{-8}$ Torr. The deposition rates of the constituent layers, calibrated with X-ray reflectivity (XRR), were 0.023 nm/s for Ti, 0.12 nm/s for Cu, 0.021 nm/s for Co, and 0.040 nm/s for Fe$_{50}$Mn$_{50}$. 
Throughout this Letter, we compare ferromagnetic Co-based film stacks grown on two types of seed layers: 
\begin{enumerate}[label=(\roman*)]
    \item 3-nm Ti (substrate/Ti/Co...) and 
    \item 3-nm-Ti/1-nm Cu (substrate/Ti/Cu/Co...).
\end{enumerate}
As we show systematically below, the 1-nm Cu seed layer has a profound impact on the structure and performance of the spin valve. Ti was selected to ensure adhesion to the oxide substrate \cite{Ohring2002,Rossnagel1994}and prevent Cu-Si intermixing~\cite{Parkin1991}. Because Ti readily oxidizes (e.g., by pulling oxygen from SiO$_{2}$), it exhibits a high electrical resistivity of $\sim 1000~\mu\Omega\cdot$cm, which ensures that the Ti seed does not act as a significant current shunt.   
Co was selected because it has the highest GMR among elemental ferromagnets. All measurements were performed in ambient air at room temperature.

To determine the quality of the interface between the seed layer and the subsequent Co layer, we conducted XRR on Ti- and Ti/Cu-seeded Co films, each $\approx$ 30 nm thick and capped with 3-nm Ti for protection against oxidation.  Figure~\ref{fig:fig1}(a) shows the XRR data, along with modeled curves produced with GenX\cite{Bjorck2007GenX} to quantify the interfacial width $\sigma$, i.e., diffuseness or roughness. (Parameters for the modeled curves are summarized in the Supplementary Material.) Of particular note is that the measured XRR intensity for Ti-seeded Co decays sharply at higher angles. This decay corresponds to a broad interfacial width of $\sigma \approx 1$ nm at the bottom Ti/Co interface, likely due to Ti-Co intermixing. By comparison, the XRR intensity for Ti/Cu-seeded Co decays much more gradually, indicating a sharp Cu/Co interface with $\sigma < 0.3$ nm. Thus, just 1 nm of Cu seed layer stabilizes a sharp bottom interface for the subsequent Co layer.

In addition to interfacial sharpness, we also anticipate the Cu seed layer to stabilize FCC (111) out-of-plane crystal orientation (texture)\cite{Liu2014_SeedLayerCu111,Joyce1998Crystallographic,Murray2006Underlayer} of the polycrystalline Co-based spin valve. To verify such Cu-aided texturing, we compare $2\theta$-$\omega$ X-ray diffraction (XRD) of Ti- and Ti/Cu-seeded Co/Cu/Co spin-valve-like stacks, each capped with Ti. Figure~\ref{fig:fig1}(b) shows XRD results for stacks with 3.0-nm Co layers and a 3.5-nm Cu spacer. The Ti-seeded stack shows only a small diffraction peak centered around $2\theta \approx 44 ^\circ$. No significant out-of-plane texturing is evidenced for Co; the small diffraction peak is attributed to the weak (111) texturing of the Cu spacer, as identical XRD is obtained from a control 3.5-nm Cu film sandwiched between 3-nm Ti (see Supplementary Material). In contrast, the Ti/Cu-seeded stack exhibits a markedly taller FCC (111) diffraction peak. Our results corroborate earlier studies \cite{489821,You2010CPP,Lenssen1997Inverted} that the Cu seed layer strongly promotes (111)-oriented growth of the subsequent FCC metal layers.

 \begin{figure*}[t] 
    \centering
     \includegraphics[width=1\textwidth]{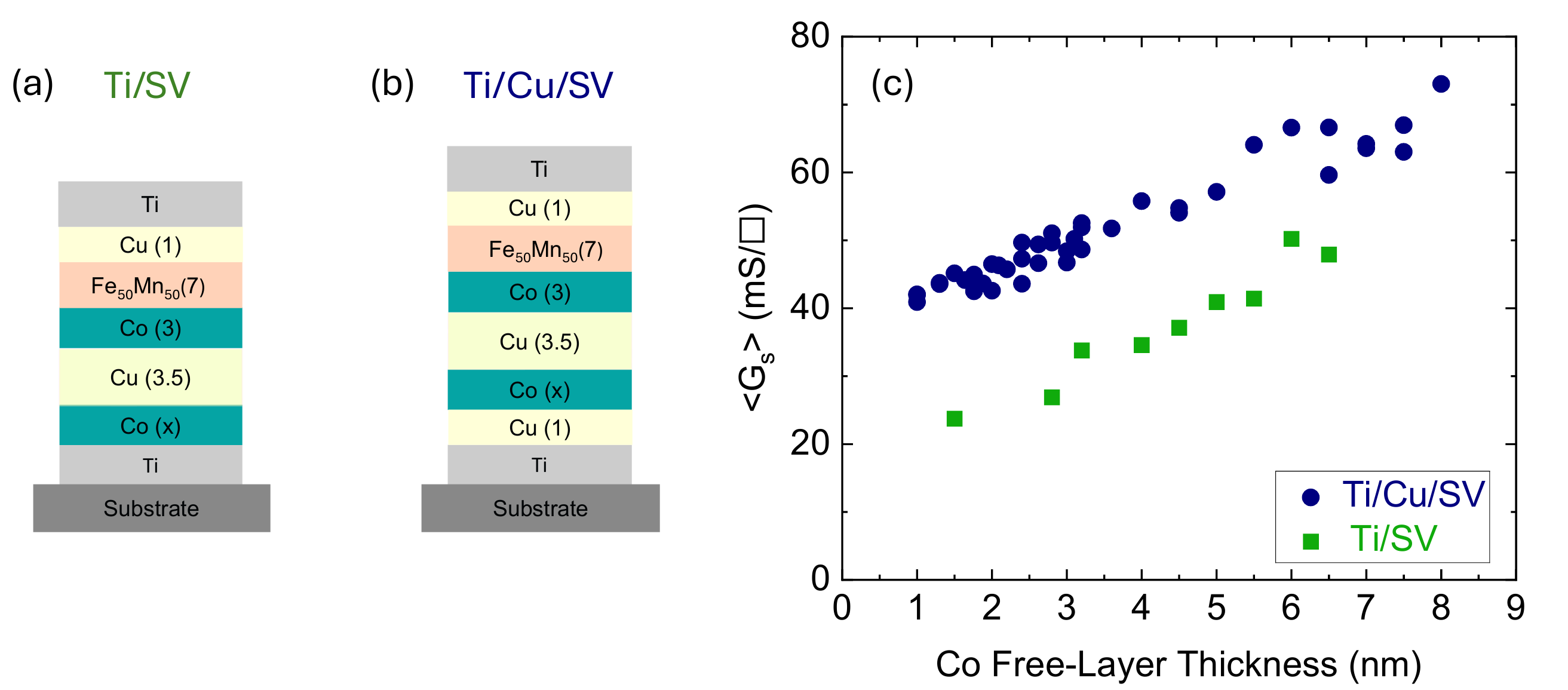} 
    \caption{\raggedright  (a,b) Schematic illustrations of the spin-valve layer stack (a) Ti/SV and (b) Ti/Cu/SV samples, showing the layer sequence and thicknesses. (c) Sheet conductance for Ti/SV and Ti/Cu/SV plotted against the Co free-layer thickness.}
    \label{fig:fig2}
\end{figure*}

We now proceed to address the crucial question: how does the inclusion of the 1-nm Cu seed layer impact the conductance and GMR of spin valves? Hereafter, each spin valve consists of \begin{quote}
substrate/Ti(3)/Cu(0 or 1)/Co($x$)/Cu(3.5)/ Co(3)/Fe$_{50}$Mn$_{50}$(7)/Cu(1)/Ti(3),\end{quote} where the values in the parentheses are layer thicknesses in nm and $x$ is the free-layer thickness. We denote Ti/Cu(0)-seeded spin valves as ``Ti/SV'' and Ti/Cu(1)-seeded spin valves as ``Ti/Cu/SV.'' Each stack has the free layer on the bottom for direct seed-layer-templated growth. Additionally, having the Co fixed layer and antiferromagnetic Fe$_{50}$Mn$_{50}$ on top allows for exchange-biasing without post-annealing.  In particular, to establish the exchange-bias direction in Co/Fe$_{50}$Mn$_{50}$, each spin valve was grown under an \emph{in-situ} forming field of 28 mT from a pair of Alnico permanent magnets affixed to the substrate stage. 

\begin{figure*}[t] 
    \centering
     \includegraphics[width=1\textwidth]{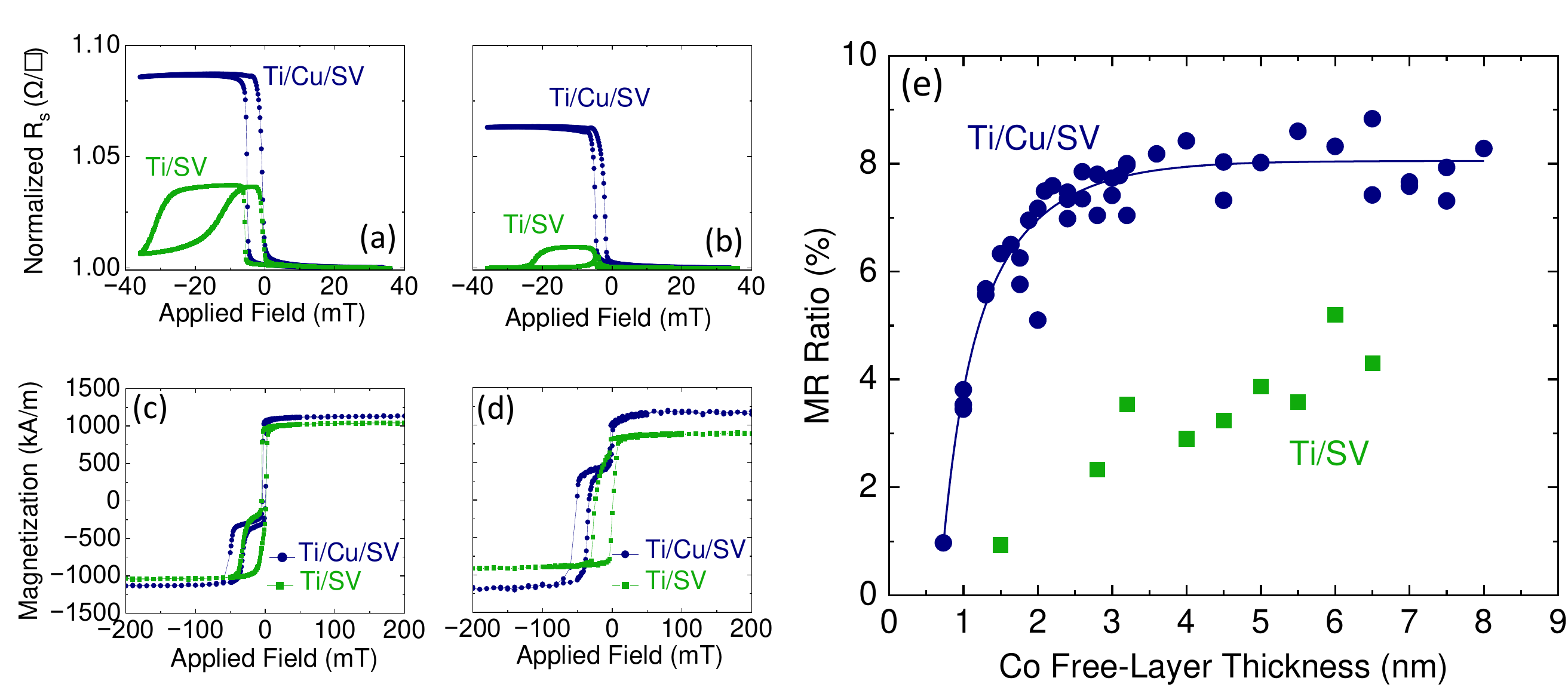} 
    \caption{\raggedright (a,b) Representative GMR curves from the Ti/SV and Ti/Cu/SV series with free-layer thicknesses of (a) $5.5$ nm and (b) $1.5$ nm. (c,d) Room-temperature magnetization curves from the Ti/SV and Ti/Cu/SV series with Co free-layer thicknesses of (c) 5.5 nm and (d) 1.5 nm. (e) GMR ratio as a function of Co free-layer thickness for Ti/SV and Ti/Cu/SV.}
    \label{fig:fig3}
\end{figure*}

Figure~\ref{fig:fig2} compares the sheet conductance, measured with the standard van der Pauw method, of the Ti/SV and Ti/Cu/SV series plotted against the Co free-layer thickness. Ti/Cu/SV shows a systematically higher sheet conductance compared to Ti/SV by up to a factor of 2. One might expect Ti/Cu/SV to be more conductive simply because the 1-nm Cu seed layer provides an additional conductive path. However, separate measurements reveal that 1-nm-thick Cu (sandwiched between 3-nm Ti layers) only has a sheet conductance of $<$ 3 mS/$\Box$. Therefore, the $\approx$20 mS/$\Box$ higher sheet conductance of Ti/Cu/SV requires another explanation.

A plausible possibility is that the superior crystalline and interfacial quality of Ti/Cu/SV leads to higher conductance than Ti/SV. In particular, our XRD results suggest about a factor of 2 larger grain size for Ti/Cu/SV (e.g., Fig.~\ref{fig:fig1}(b) $\approx6$ nm for the Ti/Cu-seeded sample and $\approx3$ nm for the Ti-seeded sample), estimated by applying Scherrer's formula to the width at half maximum of the diffraction peak. As such, Ti/Cu/SV may exhibit less electronic scattering at grain boundaries. Moreover, the sharp interfaces in Ti/Cu/SV likely favor specular electron scattering (preserving momentum) over diffuse scattering (destroying momentum) at the interfaces. From a practical perspective, the higher sheet conductance is beneficial for reducing power dissipation in the spin valve. Additionally, with the minimal current shunt through the Cu seed layer, a substantial fraction of the in-plane current can flow through the ferromagnetic layers, permitting robust GMR as we show in the following.

To characterize GMR, we measured the sheet resistance under a sweeping in-plane magnetic field applied along the exchange-bias axis. Figure \ref{fig:fig3}(a,b) shows examples of how the normalized sheet resistance evolves with the magnetic field. The hysteresis loop at low fields in each case captures the switching of the free layer. In particular, the lower resistance level $R_\mathrm{min}$ (normalized to 1 in Fig.~\ref{fig:fig3}(a,b)) indicates that the free-layer magnetization is parallel to the fixed-layer magnetization, whereas the higher resistance $R_\mathrm{max}$ indicates that the free-layer magnetization is antiparallel to the fixed-layer magnetization. We quantify the GMR ratio as 
\begin{equation}
MR = \frac{R_\mathrm{max}-R_\mathrm{min}}{R_\mathrm{min}}. 
\end{equation}

At the Co free-layer thickness of 5.5 nm, $MR \approx 8 \%$ for Ti/Cu/SV is about a factor of 2 greater than $MR \approx 4\%$ for Ti/SV [Fig.~\ref{fig:fig3}(a)]. With the free-layer thickness reduced to 1.5 nm, this difference increases to about a factor of 6, as Ti/Cu/SV maintains a high GMR ratio of $MR \approx 6\%$ while Ti/SV shows a deteriorated ratio of only $MR \approx 1\%$ [Fig.~\ref{fig:fig3}(b)]. 
We also observe that in the Ti/SV samples, the resistance reverts to $R_\mathrm{min}$ even under a modest negative applied field, indicating that the fixed-layer magnetization readily flips. Indeed, our vibrating sample magnetometry results [Fig.~\ref{fig:fig3}(c,d)] confirm that exchange-bias pinning is weak in the fixed layer of Ti/SV. The weak exchange bias is consistent with the limited crystal texturing of Ti/SV, since exchange bias in FCC systems is known to be correlated with (111) texturing~\cite{Choe1997,Castro2013}. The lack of crystal texture in the bottom Co free layer of Ti/SV propagates through the stack, producing poor texturing in the top Co fixed layer that weakens the exchange bias.

Figure \ref{fig:fig3}(c) further highlights the superiority of Ti/Cu/SV over Ti/SV, especially at smaller free-layer thicknesses. 
In the Ti/SV series, the GMR ratio dips to $MR < 2\%$ at 3 nm of free-layer thickness and essentially vanishes below 2 nm -- comparable to conventional spin valves from earlier experiments \cite{Chan2016UltrathinSV,Romera2009UltrathinGd,Dieny1991}. For instance, our results for Ti/SV quantitatively align with room-temperature GMR ratios for Co-based spin valves in Ref.~\citen{Dieny1991}, grown directly on glass substrates with no seed layers. 
In contrast, the GMR ratio of the Ti/Cu/SV series remains close to the saturated value down to a free-layer thickness of around 2 nm. Even in the free-layer thickness window of $1.3-2$ nm, we still observe high GMR ratios of $MR \approx 5-7\%$, on par with spin valves with considerably thicker free layers~\cite{Ustinov2012HysteresisFree,Dieny1991,PhysRevLett.71.1641,Kim2019TailoringGMR}. 

The stark contrast in GMR emerges from the different qualities of Ti- and Ti/Cu-seeded Co. First, the larger grain size in Ti/Cu/SV may enhance GMR by reducing spin-flip scattering at grain boundaries. More importantly, the interfacial quality of the Co free layer plays a critical role. In Ti/SV, the first few atomic monolayers of Co on Ti develop a diffuse interface (magnetic dead layer), which diminishes both the active ferromagnetic thickness and the current spin polarization. This dead-layer effect is evidenced by our magnetometry results [Fig.~\ref{fig:fig3}(c,d)], which show a systematically lower saturation magnetization for Ti/SV compared to Ti/Cu/SV. Conversely, templating Co growth with a Cu seed layer yields sharp interfaces, which correlate with enhanced GMR\cite{Kanai1993,489821,bouziane2006buffer}.
Remarkably, a mere 1-nm Cu seed layer enables the Ti/Cu/SV with a 1.5-nm free layer [Fig.~\ref{fig:fig3}(d)] to maintain a high saturation magnetization, preserving the large current spin polarization required for GMR.
We also find that increasing the Cu seed thickness yields no further GMR enhancement (see Supplementary Material). Thus, we conclude that a 1-nm Cu seed layer is sufficient to attain the interfacial sharpness and robust magnetism required for high GMR with $ \lesssim 2 $-nm free layers.

Lastly, we comment on the potential applicability and limitations of our present study. While this study focuses on elemental FCC Co, an FCC alloy of Co$_{90}$Fe$_{10}$ is often preferred for most spin valves due to its higher GMR ratio \cite{Nishioka1996GMRCoFe,Kamiguchi1996GMRCoFeCu,Yuasa2004CPPCoFe}. It is likely that the 1-nm Cu seed layer will enable elevated GMR with $\lesssim2$-nm-thin Co$_{90}$Fe$_{10}$ free layers. A similar seed-layering strategy could also be useful for GMR spin valves incorporating other FCC alloys with soft magnetism (e.g., NiFe) or thin FCC Co/Ni multilayers with perpendicular magnetic anisotropy. On the other hand, the anti-damping torque from out-of-plane spin-current in a spin valve -- necessary for robust perpendicular switching~\cite{Baek2018} and large-angle precession~\cite{PhysRevB.111.054425} -- remains to be fully understood and engineered. Nevertheless, we anticipate that the high GMR demonstrated here with ultrathin seed and free layers will greatly aid future studies of out-of-plane spin-current torques in scalable spin valves. Furthermore, while the free-layer and Cu-seed-layer thicknesses are the only variables in our present study, careful optimization of the growth condition \cite{Egelhoff1996}, improved layer continuity via the use of surfactants \cite{Egelhoff1996PbAu,Egelhoff1996GMR}, and incorporation of specular-reflection layers \cite{Hong2003,Swagten1997} may further increase the GMR ratio to over 10\%.

In summary, we have investigated current-in-plane spin valves whose free layers are seeded by Ti or Ti/Cu. Our results show that the incorporation of 1-nm Cu significantly improves the structural quality of the magnetic layers, accompanied by high GMR. This effect is especially pronounced for thinner free layers; at free-layer thicknesses of $\approx1.3-2$ nm, Cu-seeded spin valves exhibit GMR ratios of $\approx5-7$\%, in contrast to vanishing GMR for spin valves lacking a Cu seed. The high GMR ratios with such ultrathin seed and free layers are particularly attractive for spin-orbit-torque applications of spin valves. Our approach offers a scalable, effective pathway to high-signal GMR readout for emerging spintronic memories and neuromorphic computers. 

\begin{acknowledgments}
S.A. and S.E. were supported by the National Science Foundation (NSF) under Grant No. ECCS-2236160. R.E.M. was supported by the National Science Foundation (NSF) under Grant No. DMR-2144333. D.L.H. was supported by the Luther and Alice Hamlett Undergraduate Research Support program. This research used resources in the Structural Characterization Lab in the Department of Materials Science and Engineering at Virginia Tech.  
\end{acknowledgments}

\section*{Data Availability}
The data that support the findings of this study are available from the corresponding authors upon reasonable request.

\bibliography{aipsamp}

@article{RevModPhys.91.035004,
  title = {Current-induced spin-orbit torques in ferromagnetic and antiferromagnetic systems},
  author = {Manchon, A. and \ifmmode \check{Z}\else \v{Z}\fi{}elezn\'y, J. and Miron, I. M. and Jungwirth, T. and Sinova, J. and Thiaville, A. and Garello, K. and Gambardella, P.},
  journal = {Rev. Mod. Phys.},
  volume = {91},
  issue = {3},
  pages = {035004},
  numpages = {80},
  year = {2019},
  month = {Sep},
  publisher = {American Physical Society},
  doi = {10.1103/RevModPhys.91.035004},
  url = {https://link.aps.org/doi/10.1103/RevModPhys.91.035004}
}

@ARTICLE{9427163,

  author={Shao, Qiming and Li, Peng and Liu, Luqiao and Yang, Hyunsoo and Fukami, Shunsuke and Razavi, Armin and Wu, Hao and Wang, Kang and Freimuth, Frank and Mokrousov, Yuriy and Stiles, Mark D. and Emori, Satoru and Hoffmann, Axel and Åkerman, Johan and Roy, Kaushik and Wang, Jian-Ping and Yang, See-Hun and Garello, Kevin and Zhang, Wei},
  journal={IEEE Transactions on Magnetics}, 
  title={Roadmap of Spin–Orbit Torques}, 
  year={2021},
  volume={57},
  number={7},
  pages={1-39},
  doi={10.1109/TMAG.2021.3078583}
}

@inproceedings{Ustinov2012HysteresisFree,
  author    = {V. V. Ustinov and M. A. Milyaev and L. I. Naumova and V. V. Proglyado and T. P. Krinitsina},
  title     = {Hysteresis-Free Spin Valves for GMR Sensors},
  booktitle = {NSTI-Nanotech 2012 Conference Proceedings},
  series    = {TechConnect Briefs},
  year      = {2012},
  volume    = {1},
  pages     = {745},
  url       = {https://briefs.techconnect.org/wp-content/volumes/Nanotech2012v1/pdf/581.pdf},
  note      = {Paper No. 581}
}

@article{Chan2016UltrathinSV,
  title = {Spin valves with Conetic based synthetic ferrimagnet free layer},
  author = {Chan, P. H. and Li, X. and Pong, Philip W. T.},
  journal = {Vacuum},
  volume = {140},
  pages = {111--115},
  year = {2017},
  doi = {10.1016/j.vacuum.2016.09.010},
  url = {https://doi.org/10.1016/j.vacuum.2016.09.010}
}

@article{Montoya2023,
  author  = {Montoya, Eric Arturo and Khan, Amanatullah and Safranski, Christopher and Smith, Andrew and Krivorotov, Ilya N.},
  title   = {Easy-plane spin Hall oscillator},
  journal = {Communications Physics},
  volume  = {6},
  pages   = {184},
  year    = {2023},
  doi     = {10.1038/s42005-023-01298-7},
  url     = {https://doi.org/10.1038/s42005-023-01298-7},
  publisher = {Nature Publishing Group}
}

@article{Divinskiy2019,
  author  = {Boris Divinskiy and Sergei Urazhdin and Sergej O. Demokritov and Vladislav E. Demidov},
  title   = {Controlled nonlinear magnetic damping in spin-Hall nano-devices},
  journal = {Nature Communications},
  volume  = {10},
  pages   = {5211},
  year    = {2019},
  doi     = {10.1038/s41467-019-13246-7},
  url     = {https://doi.org/10.1038/s41467-019-13246-7}
}

@article{Romera2009UltrathinGd,
  title = {Influence on the magnetoresistance of a spin valve due to the insertion of an ultrathin Gd layer in the free layer},
  author = {Romera, Miguel and Mu{\~n}oz, Manuel and S{\'a}nchez, P. and Aroca, Claudio},
  journal = {Journal of Applied Physics},
  volume = {106},
  number = {2},
  year = {2009},
  doi = {10.1063/1.3173580}
}

@article{Egelhoff1996,
  author = {W. F. Egelhoff and P. J. Chen and C. J. Powell and M. D. Stiles and R. D. McMichael and C.-L. Lin and J. M. Sivertsen and J. H. Judy and K. Takano and A. E. Berkowitz and T. C. Anthony and J. A. Brug},
  title = {Optimizing the giant magnetoresistance of symmetric and bottom spin valves (invited)},
  journal = {Journal of Applied Physics},
  volume = {79},
  pages = {5277--5281},
  year = {1996},
  doi = {10.1063/1.361352},
  url = {https://doi.org/10.1063/1.361352}
}

@article{DAVIDSON2020126228,
title = {Perspectives of electrically generated spin currents in ferromagnetic materials},
journal = {Physics Letters A},
volume = {384},
number = {11},
pages = {126228},
year = {2020},
issn = {0375-9601},
doi = {https://doi.org/10.1016/j.physleta.2019.126228},
url = {https://www.sciencedirect.com/science/article/pii/S0375960119311703},
author = {Angie Davidson and Vivek P. Amin and Wafa S. Aljuaid and Paul M. Haney and Xin Fan}
}

@article{Kim2024,
  title = {Spin current and spin-orbit torque induced by ferromagnets},
  author = {Kim, Kyoung Whan and Park, Byong Guk and Lee, Kyung Jin},
  journal = {npj Spintronics},
  volume = {2},
  number = {1},
  pages = {Article 8},
  year = {2024},
  month = {Dec},
  doi = {10.1038/s44306-024-00010-x}
}

@article{Parkin1991,
  author  = {S. S. P. Parkin and Z. G. Li and David J. Smith},
  title   = {Giant magnetoresistance in antiferromagnetic Co/Cu multilayers},
  journal = {Applied Physics Letters},
  volume  = {58},
  number  = {23},
  pages   = {2710--2712},
  year    = {1991},
  doi     = {10.1063/1.104765}
}

@article{Egelhoff1992CuBuffer,
  author  = {Egelhoff, W. F. and McMichael, R. D. and Lin, C.-L. and Judy, J. H. and Takano, K. and Berkowitz, A. E.},
  title   = {Giant magnetoresistance in Co/Cu, Co9Fe/Cu, and Co7.5Fe2.5/Cu multilayers},
  journal = {IEEE Transactions on Magnetics},
  volume  = {28},
  number  = {5},
  pages   = {2742--2744},
  year    = {1992},
  doi     = {10.1109/20.179617}
}

@article{Kanai1993,
  author  = {Hitoshi Kanai and Robert L. White},
  title   = {Magnetoresistance and Structural Properties of CoFe/Cu Multilayers},
  journal = {IEEE Transactions on Magnetics},
  volume  = {29},
  number  = {6},
  pages   = {2814--2816},
  year    = {1993},
  doi     = {10.1109/20.281121}
}

@ARTICLE{489821,
  author={Gangopadhyay, S. and Shen, J.X. and Kief, M.T. and Barnard, J.A. and Parker, M.R.},
  journal={IEEE Transactions on Magnetics}, 
  title={Giant magnetoresistance in CoFe/Cu multilayers with different buffer layers and substrates}, 
  year={1995},
  volume={31},
  number={6},
  pages={3933-3935},
  doi={10.1109/20.489821}}

@article{chihaya2004effect,
  author  = {H. Chihaya and M. Kamiko and S.-M. Oh and R. Yamamoto},
  title   = {Effect of seed layers on the structure of Co/Cu (100) metallic multilayers},
  journal = {Journal of Magnetism and Magnetic Materials},
  volume  = {272--276},
  pages   = {1228--1230},
  year    = {2004},
  doi     = {10.1016/j.jmmm.2003.12.124},
  url     = {https://doi.org/10.1016/j.jmmm.2003.12.124}
}

@article{You2010CPP,
  author  = {C. Y. You and N. Tian and H. S. Goripati and T. Furubayashi},
  title   = {Current-perpendicular-to-the-plane giant magnetoresistance of an all-metal spin valve structure with Co40Fe40B20 magnetic layer},
  journal = {Applied Physics Letters},
  volume  = {96},
  number  = {14},
  pages   = {142503},
  year    = {2010},
  doi     = {10.1063/1.3385730},
  url     = {https://doi.org/10.1063/1.3385730}
}

@article{Lenssen1997Inverted,
  author  = {K.-M. H. Lenssen and A. E. M. De Veirman and J. J. T. M. Donkers},
  title   = {Inverted spin valves for magnetic heads and sensors},
  journal = {Journal of Applied Physics},
  volume  = {81},
  number  = {8},
  pages   = {4915--4917},
  year    = {1997},
  doi     = {10.1063/1.364818},
  url     = {https://doi.org/10.1063/1.364818}
}

@article{bouziane2006buffer,
  author  = {K. Bouziane and A. D. Al Rawas and M. Maaza and M. Mamor},
  title   = {Buffer effect on GMR in thin Co/Cu multilayers},
  journal = {Journal of Alloys and Compounds},
  volume  = {414},
  number  = {1-2},
  pages   = {42--47},
  year    = {2006},
  doi     = {10.1016/j.jallcom.2005.07.038},
  url     = {https://doi.org/10.1016/j.jallcom.2005.07.038}
}

@article{Baek2018,
  title = {Spin currents and spin–orbit torques in ferromagnetic trilayers},
  author = {Baek, Seung-heon C. and Amin, Vivek P. and Oh, Young-Wan and Go, Gyungchoon and Lee, Seung-Jae and Lee, Geun-Hee and Kim, Kab-Jin and Stiles, M. D. and Park, Byong-Guk and Lee, Kyung-Jin},
  journal = {Nature Materials},
  volume = {17},
  number = {6},
  pages = {509--513},
  year = {2018},
  doi = {10.1038/s41563-018-0041-5}
}

@article{PhysRevB.111.054425,
  title = {Large-amplitude easy-plane spin-orbit torque oscillators driven by out-of-plane spin current: A micromagnetic study},
  author = {Kubler, Daniel and Smith, David A. and Nguyen, Tommy and Ramos-Diaz, Fernando and Emori, Satoru and Amin, Vivek P.},
  journal = {Phys. Rev. B},
  volume = {111},
  issue = {5},
  pages = {054425},
  numpages = {12},
  year = {2025},
  month = {Feb},
  publisher = {American Physical Society},
  doi = {10.1103/PhysRevB.111.054425},
  url = {https://link.aps.org/doi/10.1103/PhysRevB.111.054425}
}

@article{Grollier2020,
  title  = {Neuromorphic spintronics},
  author = {Grollier, J. and Querlioz, D. and Camsari, K. Y. and Everschor-Sitte, K. and Fukami, S. and Stiles, M. D.},
  journal = {Nature Electronics},
  volume = {3},
  number = {7},
  pages = {360--370},
  year = {2020},
  month = {Mar},
  doi = {10.1038/s41928-019-0360-9},
  url = {https://doi.org/10.1038/s41928-019-0360-9}
}

@article{Chen2020,
  title = {Spin–orbit torque nano-oscillator with giant magnetoresistance readout},
  author = {Chen, Jen-Ru and Smith, Andrew and Montoya, Eric A. and Lu, Jia G. and Krivorotov, Ilya N.},
  journal = {Communications Physics},
  volume = {3},
  pages = {187},
  year = {2020},
  month = {Oct},
  doi = {10.1038/s42005-020-00454-7},
  url = {https://doi.org/10.1038/s42005-020-00454-7}
}

@article{Liu2020,
  title = {Synthetic antiferromagnet-based spin Josephson oscillator},
  author = {Liu, Yizhou and Barsukov, Igor and Barlas, Yafis and Krivorotov, Ilya N. and Lake, Roger K.},
  journal = {Applied Physics Letters},
  volume = {116},
  number = {13},
  pages = {132409},
  year = {2020},
  month = {Apr},
  doi = {10.1063/5.0003477},
  url = {https://doi.org/10.1063/5.0003477}
}

@article{mi12091010,
AUTHOR = {Khunkitti, Pirat and Siritaratiwat, Apirat and Pituso, Kotchakorn},
TITLE = {Free Layer Thickness Dependence of the Stability in Co2(Mn0.6Fe0.4)Ge Heusler Based CPP-GMR Read Sensor for Areal Density of 1 Tb/in2},
JOURNAL = {Micromachines},
VOLUME = {12},
YEAR = {2021},
NUMBER = {9},
ARTICLE-NUMBER = {1010},
URL = {https://www.mdpi.com/2072-666X/12/9/1010},
PubMedID = {34577654},
ISSN = {2072-666X},
DOI = {10.3390/mi12091010}
}

@article{Elsafi_2025,
doi = {10.1149/2162-8777/ade9ef},
url = {https://doi.org/10.1149/2162-8777/ade9ef},
year = {2025},
month = {jul},
publisher = {IOP Publishing},
volume = {14},
number = {7},
pages = {073002},
author = {Elsafi, Bassem},
title = {Structural and Interfacial Effects on Magnetoresistance in Ultra-Thin Co/Cu Multilayers},
journal = {ECS Journal of Solid State Science and Technology},
}

@article{parkin1990,
author    = {S. S. P. Parkin and N. More and K. P. Roche},
title     = {Oscillations in exchange coupling and magnetoresistance in metallic superlattice structures: Co/Ru, Co/Cr, and Co/Cu},
journal   = {Phys. Rev. Lett.},
volume    = {64},
number    = {19},
pages     = {2304--2307},
year      = {1990},
doi       = {10.1103/PhysRevLett.64.2304}
}

@article{PhysRevB.79.174421,
  title = {Giant magnetoresistance in electrodeposited Co-Cu/Cu multilayers: Origin of the absence of oscillatory behavior},
  author = {Bakonyi, I. and Simon, E. and T\'oth, B. G. and P\'eter, L. and Kiss, L. F.},
  journal = {Phys. Rev. B},
  volume = {79},
  issue = {17},
  pages = {174421},
  numpages = {13},
  year = {2009},
  month = {May},
  publisher = {American Physical Society},
  doi = {10.1103/PhysRevB.79.174421},
  url = {https://link.aps.org/doi/10.1103/PhysRevB.79.174421}
}

@article{parkin_oscillations_1990,
	title = {Oscillations in exchange coupling and magnetoresistance in metallic superlattice structures: {Co}/{Ru}, {Co}/{Cr}, and {Fe}/{Cr}},
	volume = {64},
	copyright = {http://link.aps.org/licenses/aps-default-license},
	issn = {0031-9007},
	shorttitle = {Oscillations in exchange coupling and magnetoresistance in metallic superlattice structures},
	url = {https://link.aps.org/doi/10.1103/PhysRevLett.64.2304},
	doi = {10.1103/PhysRevLett.64.2304},
	language = {en},
	number = {19},
	urldate = {2025-09-09},
	journal = {Physical Review Letters},
	author = {Parkin, S. S. P. and More, N. and Roche, K. P.},
	month = may,
	year = {1990},
	pages = {2304--2307},
}

@article{Zsurzsa2022,
  author    = {S{\'a}ndor Zsurzsa and Moustafa El-Tahawy and L{\'a}szl{\'o} P{\'e}ter and L{\'a}szl{\'o} Ferenc Kiss and Jen{\H o} Gubicza and Gy{\"o}rgy Moln{\'a}r and Imre Bakonyi},
  title     = {Spacer Layer Thickness Dependence of the Giant Magnetoresistance in Electrodeposited Ni-Co/Cu Multilayers},
  journal   = {Nanomaterials},
  volume    = {12},
  number    = {23},
  pages     = {4276},
  year      = {2022},
  doi       = {10.3390/nano12234276},
  pmid      = {36500898},
  pmcid     = {PMC9739252}
}

@article{Bjorck2007GenX,
  title = {GenX: an extensible X-ray reflectivity refinement program utilizing differential evolution},
  author = {Bj{\"o}rck, Mathias and Andersson, Goeran},
  journal = {Journal of Applied Crystallography},
  volume = {40},
  number = {6},
  pages = {1174--1178},
  year = {2007},
  doi = {10.1107/S0021889807045086}
}

@article{Bass_2007,
doi = {10.1088/0953-8984/19/18/183201},
url = {https://doi.org/10.1088/0953-8984/19/18/183201},
year = {2007},
month = {apr},
publisher = {},
volume = {19},
number = {18},
pages = {183201},
author = {Bass, Jack and Pratt, William P},
title = {Spin-diffusion lengths in metals and alloys, and spin-flipping at metal/metal interfaces: an
experimentalist’s critical review},
journal = {Journal of Physics: Condensed Matter},
}

@article{Haidar2013SpinPolThin,
  title = {Thickness dependence of the degree of spin polarization of electrical current in permalloy thin films},
  author = {Haidar, Mohammad and Bailleul, Matthieu},
  journal = {Physical Review B},
  volume = {88},
  pages = {054417},
  year = {2013},
  doi = {10.1103/PhysRevB.88.054417},
  url = {https://link.aps.org/doi/10.1103/PhysRevB.88.054417}
}

@article{Erekhinsky2010Surface,
  title = {Surface enhanced spin-flip scattering in lateral spin valves},
  author = {Erekhinsky, Mikhail and Sharoni, Amos and Casanova, Fèlix and Schuller, Ivan K.},
  journal = {Applied Physics Letters},
  volume = {96},
  number = {2},
  pages = {022513},
  year = {2010},
  doi = {10.1063/1.3291047},
}

@article{Alcer2017Mesoscopic,
  title = {The role of mesoscopic structuring on the intermixing of spin‐polarised conduction channels in thin‐film ferromagnets for spintronics},
  author = {Alcer, D. and Atkinson, D.},
  journal = {Nanotechnology},
  volume = {28},
  number = {37},
  pages = {375703},
  year = {2017},
  doi = {10.1088/1361-6528/aa7dcb},
  url = {https://iopscience.iop.org/article/10.1088/1361-6528/aa7dcb}
}

@article{Cormier2010DomainWalls,
  title = {Effect of electrical current pulses on domain walls in Pt/Co/Pt nanotracks with out-of-plane anisotropy: Spin transfer torque versus Joule heating},
  author = {Cormier, M. and Mougin, A. and Ferr\'e, J. and Thiaville, A. and Charpentier, N. and Pi\'echon, F. and Weil, R. and Baltz, V. and Rodmacq, B.},
  journal = {Physical Review B},
  volume = {81},
  pages = {024407},
  year = {2010},
  doi = {10.1103/PhysRevB.81.024407},
  url = {https://doi.org/10.1103/PhysRevB.81.024407}
}

@article{Dieny1991,
  author  = {Dieny, B. and Humbert, P. and Speriosu, V. S. and Metin, S. and Gurney, B. A. and Baumgart, P. and Lefakis, H.},
  title   = {Giant magnetoresistance of magnetically soft sandwiches: Dependence on temperature and on layer thicknesses},
  journal = {Physical Review B},
  volume  = {45},
  number  = {2},
  pages   = {806--813},
  year    = {1992},
  doi     = {10.1103/PhysRevB.45.806},
  url     = {https://doi.org/10.1103/PhysRevB.45.806}
}

@article{PhysRevLett.71.1641,
  title = {Origin of enhanced magnetoresistance of magnetic multilayers: Spin-dependent scattering from magnetic interface states},
  author = {Parkin, S. S. P.},
  journal = {Physical Review Letters},
  volume = {71},
  pages = {1641--1644},
  year = {1993},
  doi = {10.1103/PhysRevLett.71.1641},
  url = {https://doi.org/10.1103/PhysRevLett.71.1641}
}

@article{Kim2019TailoringGMR,
  title = {Tailoring of magnetic properties of giant magnetoresistance spin valves via insertion of ultrathin non-magnetic spacers between pinned and pinning layers},
  author = {Kim, Si Nyeon and Choi, Jun Woo and Lim, Sang Ho},
  journal = {Scientific Reports},
  volume = {9},
  number = {1617},
  year = {2019},
  doi = {10.1038/s41598-018-38269-w},
  url = {https://doi.org/10.1038/s41598-018-38269-w}
}

@book{Ohring2002,
  author    = {Milton Ohring},
  title     = {Materials Science of Thin Films: Deposition and Structure},
  edition   = {2},
  publisher = {Academic Press},
  address   = {San Diego, CA},
  year      = {2002},
  isbn      = {9780125249751},
  doi       = {10.1016/B978-0-12-524975-1.X5000-9},
  url       = {https://doi.org/10.1016/B978-0-12-524975-1.X5000-9}
}

@article{Rossnagel1994,
  author  = {Stephen M. Rossnagel and Jeffrey Hopwood},
  title   = {Magnetron sputter deposition with high levels of metal ionization},
  journal = {Journal of Vacuum Science \& Technology B},
  volume  = {12},
  number  = {1},
  pages   = {449--453},
  year    = {1994},
  doi     = {10.1116/1.587099}
}

@article{Liu2014_SeedLayerCu111,
  author    = {Chien-Min Liu and Han-Wen Lin and Chia-Ling Lu and Chih Chen},
  title     = {Effect of grain orientations of Cu seed layers on the growth of <111>-oriented nanotwinned Cu},
  journal   = {Scientific Reports},
  volume    = {4},
  pages     = {6123},
  year      = {2014},
  doi       = {10.1038/srep06123},
  pmid      = {25134840},
  pmcid     = {PMC4137260}
}

@article{Joyce1998Crystallographic,
  title = {Crystallographic texture and interface structure in Co/Cu multilayer films},
  author = {Joyce, D. E. and Faunce, C. A. and Grundy, P. J. and Fulthorpe, B. D. and Hase, T. P. A. and Pape, I. and Tanner, B. K.},
  journal = {Physical Review B},
  volume = {58},
  number = {9},
  pages = {5594--5601},
  year = {1998},
  doi = {10.1103/PhysRevB.58.5594},
  url = {https://doi.org/10.1103/PhysRevB.58.5594}
}

@article{Murray2006Underlayer,
  title = {Underlayer effects on texture evolution in copper films},
  author = {Murray, C. E. and Rodbell, K. P. and Vereecken, P. M.},
  journal = {Thin Solid Films},
  volume = {503},
  pages = {207--211},
  year = {2006},
  doi = {10.1016/j.tsf.2005.11.105},
  url = {https://doi.org/10.1016/j.tsf.2005.11.105}
}

@article{Choe1997,
  author = {Choe, G. and Gupta, S.},
  title = {High exchange anisotropy and high blocking temperature in strongly textured NiFe(111)/FeMn(111) films},
  journal = {Applied Physics Letters},
  volume = {70},
  number = {14},
  pages = {1766--1768},
  year = {1997},
  doi = {10.1063/1.118650}
}

@article{Castro2013,
  title = {The role of the (111) texture on the exchange bias and interlayer coupling effects observed in sputtered NiFe/IrMn/Co trilayers},
  author = {Castro, I. L. and Nascimento, V. P. and Passamani, E. C. and Takeuchi, A. Y. and Larica, C. and Tafur, M. and Pelegrini, F.},
  journal = {Journal of Applied Physics},
  volume = {113},
  number = {20},
  pages = {203903},
  year = {2013},
  doi = {10.1063/1.4804671},
  url = {https://doi.org/10.1063/1.4804671}
}

@article{Nishioka1996GMRCoFe,
  author  = {K. Nishioka and T. Iseki and M. A. Parker},
  title   = {GMR properties of spin valves using multilayered Co$_{90}$Fe$_{10}$ for free magnetic layer},
  journal = {Journal of Applied Physics},
  volume  = {79},
  number  = {8},
  pages   = {6378--6380},
  year    = {1996},
  doi     = {10.1063/1.362189}
}

@article{Kamiguchi1996GMRCoFeCu,
  author  = {Y. Kamiguchi and K. Saito and H. Iwasaki and M. Sahashi and M. Ouse and S. Nakamura},
  title   = {Giant magnetoresistance and soft magnetic properties of Co$_{90}$Fe$_{10}$/Cu spin‑valve structures},
  journal = {Journal of Applied Physics},
  volume  = {79},
  number  = {10},
  pages   = {6399--6401},
  year    = {1996},
  doi     = {10.1063/1.362011}
}

@article{Yuasa2004CPPCoFe,
  author  = {H. Yuasa and H. Fukuzawa and H. Iwasaki},
  title   = {CPP–GMR of spin valves with Co$_x$Fe$_{1-x}$ alloy},
  journal = {Journal of Magnetism and Magnetic Materials},
  volume  = {286},
  pages   = {95--98},
  year    = {2005},
  doi     = {10.1016/j.jmmm.2004.09.045}
}

\end{document}